\documentclass[%
 aip,jcp,reprint
]{revtex4-1}

\usepackage{graphicx}
\usepackage{dcolumn}
\usepackage{bm}
\usepackage{hyperref}
\usepackage{array} 
\usepackage[export]{adjustbox}
\usepackage[nomessages]{fp}
\usepackage{xcolor}
\usepackage{placeins}
\usepackage[version=3]{mhchem} 
\usepackage[T1]{fontenc}       
\usepackage{textgreek}
\usepackage{amsmath}
\usepackage[activate={true,nocompatibility},kerning=true,factor=1100,stretch=10,shrink=10]{microtype}

\begin{document}

\author{John E. Herr}
\email{jherr1@nd.edu}
\author{Kevin Koh}
\author{Kun Yao}
\author{John Parkhill}
\affiliation{%
 Dept. of Chemistry and Biochemistry, The University of Notre Dame du Lac 
}%

\date{\today}

\title{Compressing physical properties of atomic species for improving predictive chemistry.}

\begin{abstract}
The answers to many unsolved problems lie in the intractable chemical space of molecules and materials. Machine learning techniques are rapidly growing in popularity as a way to compress and explore chemical space efficiently. One of the most important aspects of machine learning techniques is representation through the feature vector, which should contain the most important descriptors necessary to make accurate predictions, not least of which is the atomic species in the molecule or material. In this work we introduce a compressed representation of physical properties for atomic species we call the elemental modes. The elemental modes provide an excellent representation by capturing many of the nuances of the periodic table and the similarity of atomic species. We apply the elemental modes to several different tasks for machine learning algorithms and show that they enable us to make improvements to these tasks even beyond simply achieving higher accuracy predictions.
\end{abstract}

\maketitle

\section{\label{sec:level1}Introduction}
\indent Machine learning is being applied at an unprecedented rate to efficiently explore the vastness of chemical space. Researchers have used a wide array of machine learning techniques to make reaction outcome predictions,\cite{schwaller2018found,nam2016linking,liu2017retrosynthetic} reaction yield predictions,\cite{nielsen2018deoxyfluorination,ahneman2018predicting} to predict bond energies,\cite{yao2017intrinsic} partial charges,\cite{nebgen2018transferable,sifain2018discovering} formation energies,\cite{faber2016machine,zhou2018learning} among other properties of electronic structure.\cite{montavon2013machine,kitchin2018machine,schutt2017quantum,rupp2012fast,hansen2015machine} There has also been great interest in generating machine learned model chemistries which promise to greatly reduce the cost of running quantum-accuracy simulations by compressing the corpus of completed results.\cite{yao2018tensormol,smith2017ani,behler2007generalized,behler2011neural,behler2008metadynamics,behler2007representing,yao2017many,yao2016kinetic,gastegger2016comparing,gastegger2017machine,brockherde2017bypassing,li2016understanding,Snyder:2013aa,snyder2012finding,behler2011atomcentered,handley2010potential,zhang2018deep,han2017deep,chmiela2017machine,schutt2016naturecm,schutt2017quantum} \\
\indent There are many challenges that come with the design of machine learning algorithms for predictive chemistry. They must respect physical invariances,\cite{thomas2018tensor} the predictions must fall within the scope of the underlying data,\cite{herr2018metadynamics,smith2018less} and predictions must be smooth with respect to geometrical changes in the molecule. Depending on the target problem, one particularly challenging aspect is the representation of atomic species and how predictions change with a change in atomic number. Typically this can be dealt with by parameterizing distinct machine learning models for each atomic species,\cite{behler2007representing,behler2011neural,yao2018tensormol,smith2017ani,yao2017intrinsic} or by representation in the feature vector.\cite{schutt2018schnet,schutt2017schnet,gastegger2018wacsf,de2016comparing,bartok2017machine,willatt2018data,faber2018alchemical,zhou2018learning} \\
\indent There are several issues with the former solution, including increased computational cost and memory requirements, but more importantly distinct parameterizations don't allow for similar atomic species to share information in the learning process. Using a single machine learning model for all atomic species would encourage faster training with less data. This can be thought of as transfer learning, since making predictions for different atomic species is often a highly similar task with some variation in the desired outcome.\cite{pan2010survey} \\
\indent Furthermore, often the feature vector of an atom must also represent the species of its neighboring atoms, such as is the case for the symmetry functions introduced by Behler and Parinello.\cite{behler2007generalized,behler2007representing} When limited to a single system, atomic species can often be implied since the data will always be consistent in this way, but in the more general case when we wish to make predictions for unique molecules or materials then this change must be represented through the feature vectors for the machine learning model to account for this change. In the case of the symmetry functions this has been done by splitting the features of neighboring atoms into element channels for the radial descriptor, and element-pair channels for the angular descriptor.\cite{smith2017ani,yao2018tensormol} This leads to a quadratically growing size of the feature vector which rapidly becomes unmanageable if treating more than a few unique atomic species. \\
\indent Recently there has been work focused on representation of atomic species in several machine learning applications. Naturally the first thought is to use atomic number as a multiplying factor to the features.\cite{gastegger2018wacsf} Others have sought to apply similar logic by using the group and period a species belongs to,\cite{faber2018alchemical,faber2017prediction} by using one or two physical properties,\cite{de2016comparing,bartok2017machine,willatt2018data} or by random initialization and allowing the multiplying factor to be learned during the training process.\cite{schutt2017schnet,schutt2018schnet} More recently, Zhou et al. proposed an atom vector which is derived from a materials dataset of which surrounding environments each atomic species appears in.\cite{zhou2018learning} \\
\indent In this work we introduce what we refer to as the elemental modes. The elemental modes improve upon the previous works discussed above by providing an atomic species vector which maps similar species nearby based on using many physical properties fundamental to each atom in a compressed representation. We avoid using a dataset of atomic environments as this will bias the representation towards the given dataset. We show that the elemental modes are highly generalizable by using them in distinctly different learning tasks with success.
\begin{figure*}[t]
\includegraphics[width=1.0\textwidth]{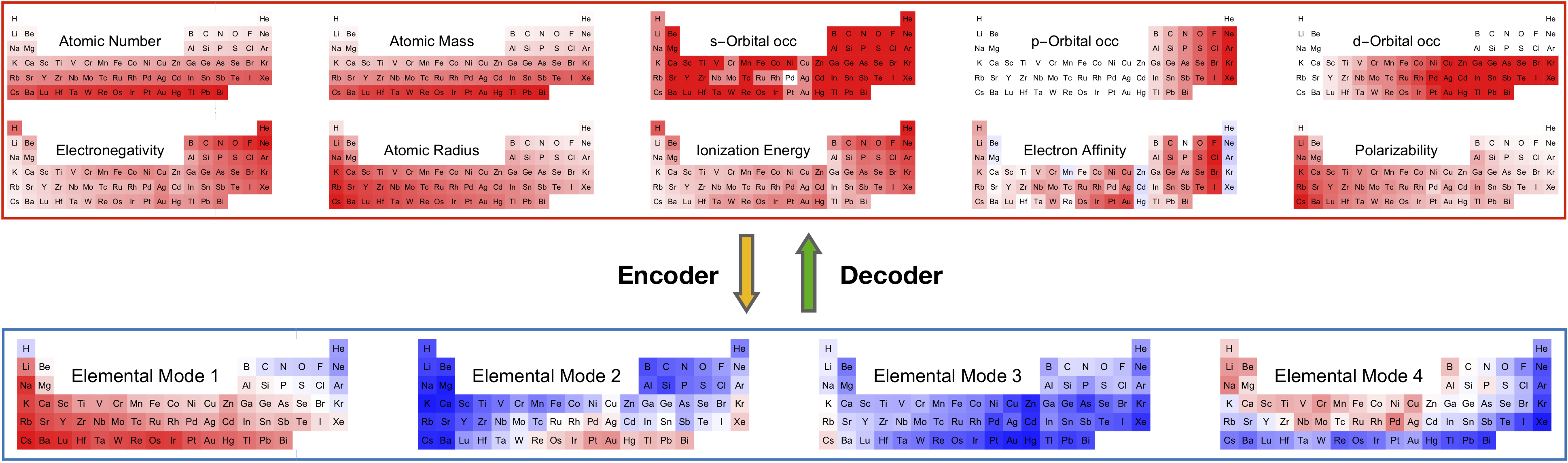}
\caption{Scheme of fundamental properties being encoded to elemental modes. 10 different properties were reduced to 6 elemental modes. Blue is negative, white is zero and red is positive.}
\label{fig:EC}
\end{figure*}
\section{Elemental modes}
\begin{figure}[t]
\includegraphics[width=1.0\columnwidth]{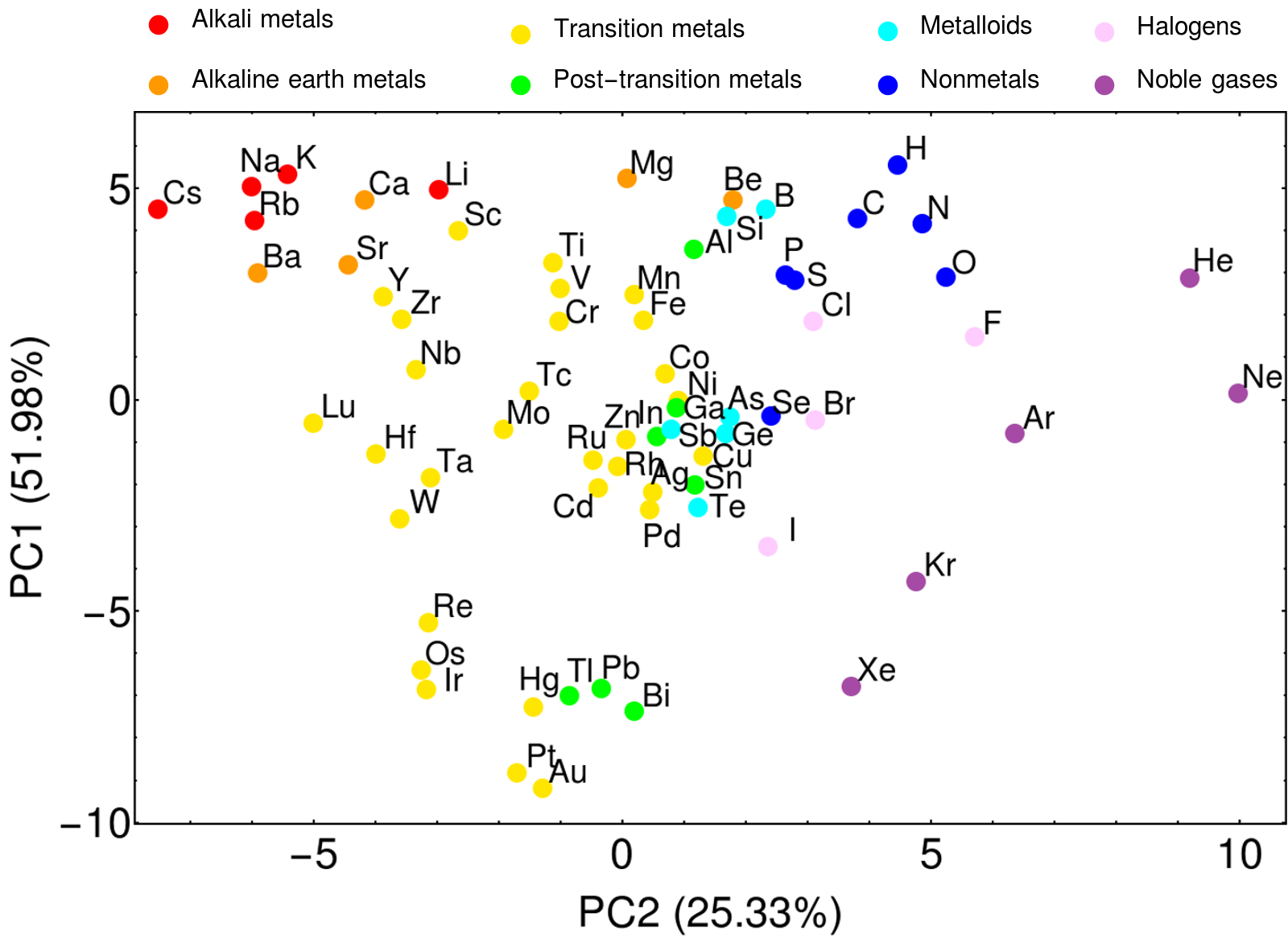}
\caption{The first (PC1) and second (PC2) principal components of the elemental modes color coded by grouping into alkali metals (red), alkaline earth metals (orange), transition metals (yellow), post-transition metals (green), metalloids (teal), nonmetals (blue), halogens (pink), and noble gases (purple).}
\label{fig:pca}
\end{figure}

\indent We compiled a dataset of fundamental physical properties for each element with atomic number up to 83 excluding f-block elements. The fundamental properties include atomic number, atomic mass, number of s, p, and d valence electrons, atomic radius, electronegativity, ionization energy, electron affinity, and polarizability. Using this data, we then trained an ordinary auto-encoder\cite{hinton2006reducing} with each element representing one case, and the ten physical properties listed above concatenated into the feature vector for each element. Figure \ref{fig:EC} shows heat map plots of the physical properties and the resulting elemental modes features as derived from the auto-encoder. \\
\indent The encoder and decoder branches each contained two hidden layers with 64 neurons in each layer. The activation function used was the hyperbolic tangent function. After training the auto-encoder to convergence, the latent space vector for each element was taken as the elemental modes. We tested different sizes of the latent space vectors, and a dimension of size four gave the best trade-off between reproduction of the physical properties from the decoder and compression of the original data. \\
\indent To examine trends learned by the auto-encoder, we performed principal component analysis on the resulting elemental modes and plot the first and second principal components in Figure \ref{fig:pca}. The trend is strikingly similar to the periodic table. Alkali metals and alkaline earth metals tend towards large positive (negative) values in the first (second) principal component. Noble gases and halogens are the opposite, tending towards large positive values in the second principal component, and large negative or smaller positive values in the first principal component. Atomic species in later periods also tend towards lower values in both the first and second principal components as opposed to those in the same group appearing in earlier periods. Having satisfied that the elemental modes do indeed represent sensible trends to encode the relationship between atomic species, we next turn our attention to their utility.
\section{Formation energy prediction}
\begin{figure}[t]
\includegraphics[width=1.0\columnwidth]{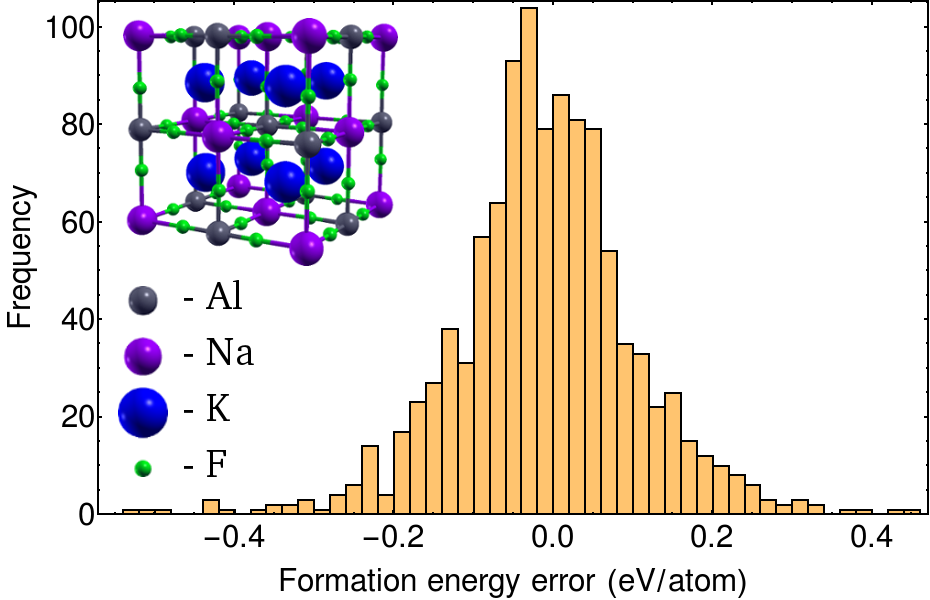}
\caption{Formation energy error prediction for the independent test set of elpasolites with the formula ABC$_{2}$D$_{6}$. Inset in the top left is the crystal structure for elpasolite (AlNaK$_{2}$F$_{6}$) }
\label{fig:elpas}
\end{figure}

\indent Recently several works have demonstrated machine learning algorithms which are capable of predicting the formation energies of elpasolites, crystals with the chemical formula ABC$_{2}$D$_{6}$.\cite{faber2016machine,zhou2018learning} The dataset used in these works consists of $\sim$ 10,000 formation energies from density functional theory (DFT) calculations of crystal structures in the elpasolite configuration containing only main group elements.\cite{faber2016machine} Restricting the data to materials of the elpasolite structure allows for the learning problem to be greatly simplified, as the periodic nature and common crystalline structure of the materials allows for a lot of information to be inherently assumed in the learning problem. As such, the most important information that varies between each structure is the atomic species corresponding to A, B, C, and D in the ABC$_{2}$D$_{6}$ formula. \\
\indent We applied the elemental modes as features for a neural network to predict formation energies on this dataset. Our feature vector is the concatenation of the elemental modes for the atomic species A, B, C, and D for a particular elpasolite structure. The feature vectors were then fed into a standard feed-forward neural network with two hidden layers of 32 neurons and a softplus activation function. \\
\indent The dataset of structures and formation energies was split into an 80:10:10 ratio for training, testing, and validation data. The network was found to perform best with two hidden layers and 32 neurons in each layer. After training had converged, the mean absolute error (MAE) on the independent test set was 0.086 eV/atom, outperforming previous works on this task. Figure \ref{fig:elpas} shows the resulting distribution of error in the predicted formation energies for the independent test set.
\section{Neural network model chemistry}
\begin{figure}
    \centering
    \includegraphics[width=0.98\columnwidth]{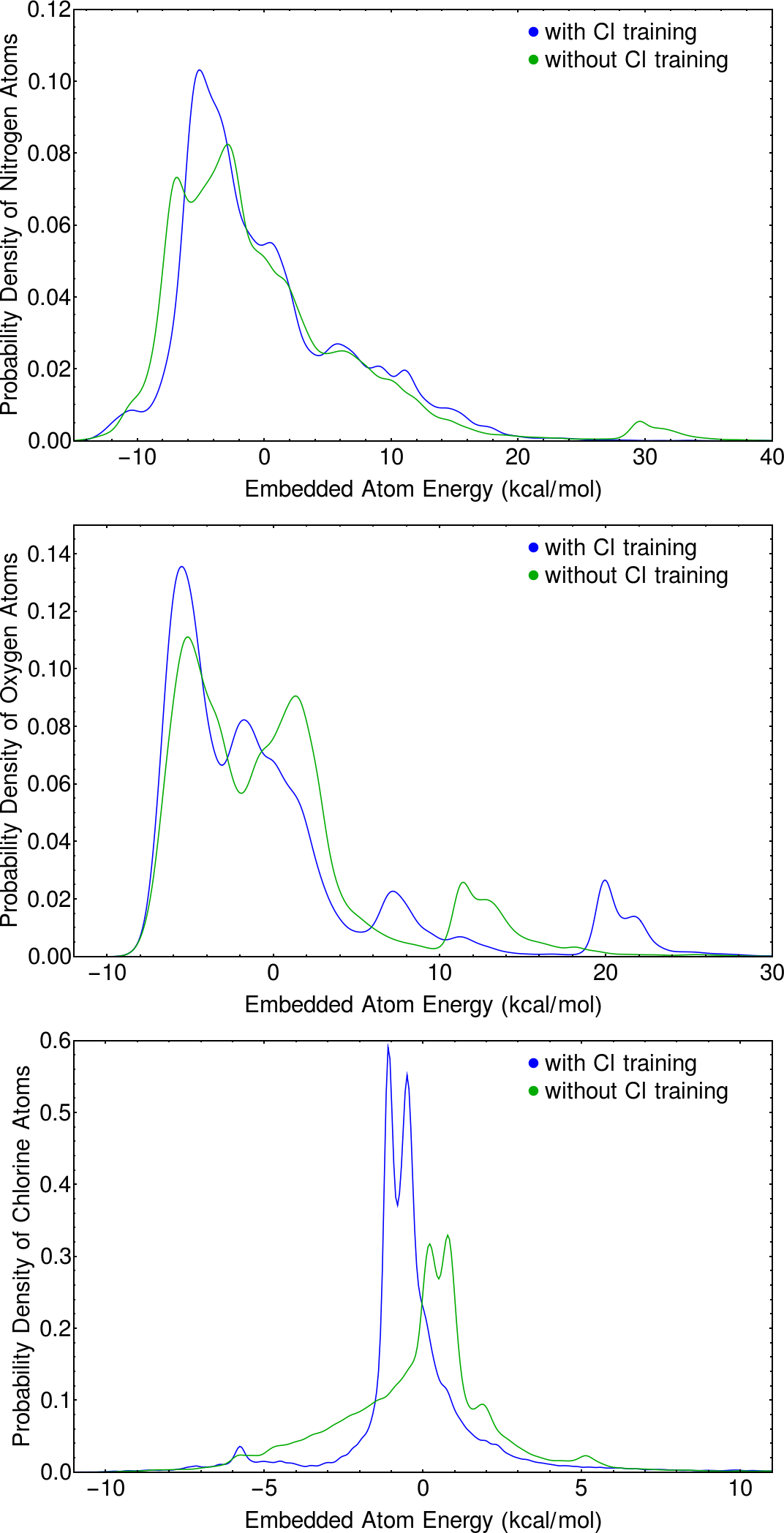}
    \caption{Probability distribution of embedded atomic energies predicted by two networks trained only on non-metals. Top, middle, and bottom panels are for nitrogen, oxygen, and chlorine predictions respectively. Distributions are included from the networks trained with (blue) and without (green) chlorine data included.}
    \label{fig:hist}
\end{figure}

\indent Next we turn our attention to the more difficult task of predicting accurate potential energies, atomic forces, and partial charges for a wide range of small molecules. The difficulty is a result of how robust the machine learning algorithm must be in this case. Unlike periodic crystal structures, there is little information which can be inherently assumed by the algorithm. In this task we must be able to represent not only the atomic species present, but also the full geometry of the molecule must be represented in the feature vector. There are several different proposed methods for making this representation,\cite{schutt2017schnet,schutt2018schnet,bartok2013representing,bartok2017machine,faber2016machine,BAML,BoBs,collins2017constant} but we will restrict this work to the symmetry functions with high-dimensional neural networks potentials (HD-NNPs) as first introduced by Behler and Parinello.\cite{behler2007generalized,behler2007representing,behler2011neural,behler2011atomcentered} \\
\indent Analogously to the work above on elpasolites where the geometry could be inherently assumed, this initial works using the symmetry functions and HD-NNPs were restricted instead to cases where the atomic species could be inherently assumed. Following researchers were able to generalize the symmetry functions to include atomic species by splitting the symmetry functions into channels for each element in the radial descriptor, and channels for each pair of elements in the angular descriptor.\cite{smith2017ani,yao2018tensormol} For a dataset containing molecules restricted to C, H, N, and O atoms, this resulted in four radial channels, and ten angular channels. Going much further in terms of atomic species allowed in the dataset will then quickly grow the size of the symmetry functions beyond what is reasonable given the memory constraints of modern graphics processing units (GPUs) needed for training these machine learning algorithms. \\
\indent To remedy this problem, we applied the elemental modes as an embedding factor into the channels of the symmetry functions to keep the size constant while allowing for any number of atomic species to be included in the dataset. More precisely, for a neighboring atom $j$ in the environment of atom $i$, the radial descriptor is given by
\begin{align}[h]
    G_{\epsilon, R_s}^R = e^{-\eta\left( R_{ij} - R_s \right)^2} f_c\left(R_{ij} \right) \otimes \beta_\epsilon(Z_j),
\end{align}
and for two neighboring atoms $j$ and $k$ in the environment of atom $i$ the angular descriptor is given by
\begin{align}
\begin{split}
    G_{\epsilon, R_s,\theta_s,}^A = 2^{1-\zeta} \left(1+cos \left(\theta_{ijk} - \theta_s \right) \right)^\zeta \\
     e^{-\eta\left( \frac{R_{ij} + R_{ik}}{2} - R_s \right)^2}
     f_c\left(R_{ij}  \right) f_c\left(R_{ik}  \right) \otimes \beta_\epsilon(Z_j) \beta_\epsilon(Z_k),
\end{split}
\end{align}
where $R_{ij}$ is the distance between atoms $i$ and $j$, $\theta_{ijk}$ is the angle between the atoms $j$, $i$, and $k$, $\beta_\epsilon(Z_j)$ is the elemental modes for the atomic species of atom $j$, $R_s$, $\theta_s$, $\zeta$, and $\eta$ are parameters of the symmetry functions, $\otimes$ represents the outer product, and $f_c\left(R_{ij} \right)$ is a smooth radial cutoff function given by
\begin{align}
    f_c(R_{ij}) = \begin{cases} 
      0.5 \times cos \left( \frac{\pi R_{ij}}{R_c} \right) + 0.5 & R_{ij} \leq R_c \\
      0 & R_{ij} > R_c 
   \end{cases}.
\end{align}
The total radial environment for atom $i$ is then given by summing over all atoms $j$ in the environment of atom $i$, and the total angular environment is given by summing over all pairs of atoms $j$ and $k$ in the environment of atom $i$. The feature vector of atom $i$ is then the concatenation of the radial and angular environments. By splitting the radial and angular environments into channels corresponding to the elemental modes, then the machine learning algorithm must infer the atomic species of atoms based on the relative scaling of values across the channels of the symmetry functions. \\
\indent Furthermore, we also wish to eliminate the need for separate neural networks to be trained for predicting the energy contribution of different atomic species. HD-NNPs make the assumption that the energy of a molecule can be broken down by summing the embedded atomic energy predicted for each atom in a molecule. Partitioning the total energy into a sum of embedded components has a history of success in the computational chemical sciences.\cite{richard2012generalized, mayhall2012many, medders2015representation, ruff2015camelot, john2017many} A choice of molecular partitioning is a compromise between two limits. If large fragments are chosen, nuanced intra-fragment physical forces are coarse-grained out of the learning problem; however, the number of such unique fragments will be large. Choosing a small fragment such as an atom requires a neural network to learn challenging physical interactions, but readily generalizes to new unseen fragments. \\
\indent An analogy can be drawn to word-level\cite{brown1992class,mikolov2013efficient,turian2010word} and character-level\cite{kim2016character,sutskever2011generating} language modeling. The meaning of a character is strongly affected by its environment, whereas a word has significant meaning on its own. We took particular note of a work from Sutskever et al. on character-level language modeling.\cite{sutskever2011generating} To account for the differences between interactions of various characters they used multiplicative interactions which allows the network weights to respond to the identity of a character embedded inside of a larger sentence. We applied similar logic by allowing our neural network to respond to the atomic species of an embedded atom. Given the feature vector of an atom $G_{i,\epsilon'}$ and the elemental modes, $\beta_{i,\epsilon''}$, corresponding to the atomic species of that atom, then we introduce to learnable matrices, ($P$,$Q$). These matrices are used to interact an atoms elemental modes with its feature vector by
\begin{align}
    I_{i,\epsilon} = P_{\epsilon',\epsilon} G_{i,\epsilon'} \beta_{i,\epsilon''} Q_{\epsilon'',\epsilon'}.
\end{align}
Allowing the feature vector of an atom to interact with it's own elemental modes then provides a way for the neural network to respond to the atomic species of an atom so that the predicted embedded atomic energy can change accordingly. \\
\indent We trained two neural networks, one which predicts embedded atomic energies and one which predicts partial charges on each atom. In a similar vein to our previous work,\cite{yao2018tensormol} the predicted partial charges were used to calculate Coulomb energies for a long-range and smoothly cutoff Coulomb kernel. As we have shown previously this helps to account for long-range interactions that the neural network cannot learn due to the short-range nature of the symmetry functions. \\
\indent Our dataset consists of about 4.3 million geometries from 65,000 unique molecules. The atomic species represented in our dataset include all nonmetals, which allows us to make predictions on a drastically more diverse set of molecules as opposed to previous works where predictions were limited to molecules containing only C, H, N, and O atoms. Initial molecular geometries were downloaded from the chemspider database and geometries were optimized to convergence. Then a subset of all 65,000 molecules was used for running metadynamics simulations to efficiently sample a more diverse set of geometries according to our previous work.\cite{herr2018metadynamics} Potential energies, atomic forces, and mulliken charges were calculated using Q-Chem\cite{shao2015advances} with the \textomega B97X-D exchange-correlation functional\cite{chai2008long} and 6-311G** basis set. \\
\indent The neural networks worked well with three hidden layers and 512 neurons in each hidden layer. The loss function used to train our network is the mean square error and included terms from the energy, atomic force, and partial charge errors. We used a softplus activation function which was chosen because it has fewer problems with vanishing gradients, similar to the ReLU or ELU activation functions, but is also continuously differentiable at least up to order two. This property is important when using atomic forces in the loss function. Since the atomic forces predicted by the neural network are the negative gradient of the predicted potential energy with respect to atomic position, then to calculate the gradients to update the network parameters will require taking second order gradients in the backpropagation algorithm. \\
\indent The root mean square error (RMSE) on the independent test set of the energy is 0.0976 kcal/mol per atom and the RMSE of the atomic forces is 3.71 kcal/mol/\AA. We note that the largest errors tend to occur in molecules which contain atoms that occur less frequently throughout the dataset. The nature of known small molecules drastically overrepresents carbon and hydrogen, and to a lesser extent nitrogen and oxygen. While our network allows to reduce this problem to an extent since a single network is used for predicting embedded atomic energies, imbalances in the dataset can still lead to larger errors for atomic species which are underrepresented. Another source of error could be the difficulty in learning the $P$ and $Q$ matrices for interacting the feature vectors with the corresponding elemental modes. As Sutskever and coworkers pointed out, learning a tensor decomposition like this is a difficult problem with first-order optimization methods alone.\cite{sutskever2011generating} Future improvements will likely focus in this direction. \\
\indent As stated above, using a single network to make the embedded atomic energy predictions should allow for the network to use information learned from one atomic species to improve the predictions of another species. To asess how well the network was able to learn accross species, we trained two more networks on subsets of the data. First we took the subset of geometries which contained at least one nonmetal other than C, H, N, and O. Then we took this same subset, but additionally removed any molecules which contained Cl atoms. We trained networks on both of these sets of data, and compared the distributions of embedded atomic energy predictions from different species, including Cl. \\
\indent Figure \ref{fig:hist} shows the atomic energy distributions from both networks for N, O and Cl for the set of Cl-containing molecules which neither network has been trained on. We note that since the embedded atomic energy was not a learning target of our neural network, then it should not be expected that that these distributions look identical, but we do notice a strong similarity for both N and O. Looking at the distribution for the Cl atomic energies, we also notice a strong similarity between the two distributions. This shows that, even though one of the networks was never trained using any molecules which contain Cl atoms, it has already reasonably learned to make sensible predictions for the embedded atomic energies of Cl atoms. \\
\section{Alchemical transformations}
\begin{figure}
    \centering
    \includegraphics[width=1.0\columnwidth]{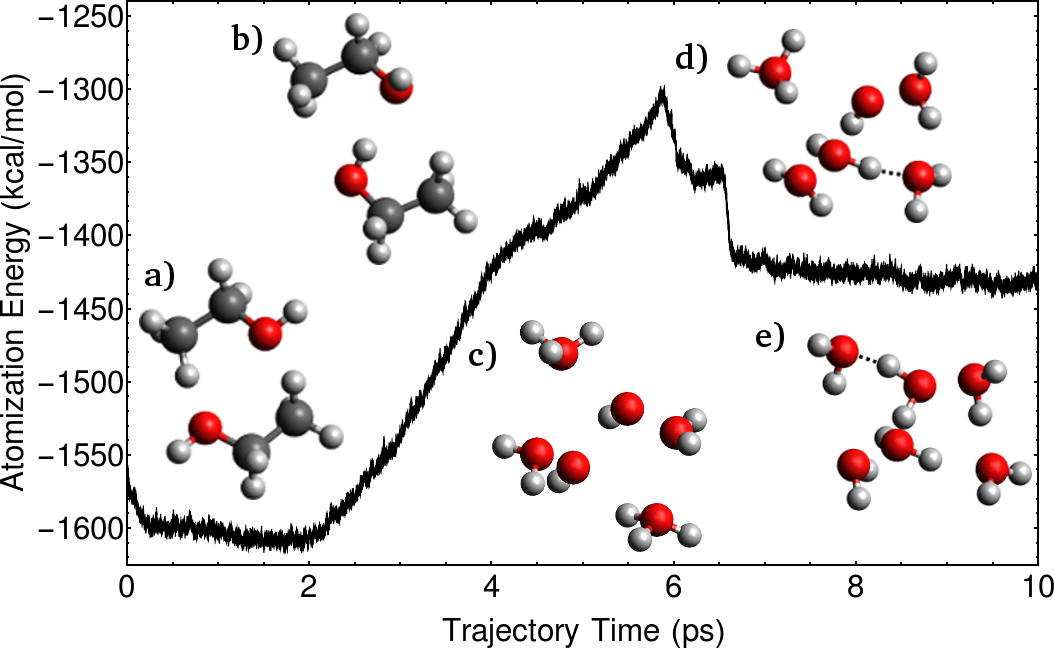}
    \caption{Plot: Atomization energy of an MD trajectory with an alchemical transformation of an ethanol dimer into a water hexamer over a 10 ps simulation time. Insets: a) The initial ethanol dimer geometry. b) The ethanol dimer geometry just before the alchemical transformation begins at 2 ps. c) The water hexamer just after the alchemical transformation completes at 5 ps. d) The first and e) second proton transfers from hydronium to hydroxide at about 6 and 6.5 ps. Proton transfer is denoted with a dashed line.}
    \label{fig:alchem}
\end{figure}

\indent Another advantage of our neural network model chemistry allows us to interpolate between the elemental modes of two atomic species and make "alchemical" energy predictions. Alchemical free energy calculations are an important tool for pharmaceutical drug discovery used by many researchers today.\cite{hauser2018predicting,mey2018impact,harger2017tinker,williams2017free,matricon2017fragment} These calculations are often difficult to set up, taking great care in switching on and off of force field parameters. Further, many molecular dynamics software packages do not include support for alchemical free energies. We have implemented the ability for our network to make potential energy predictions for the intermediate states of an alchemical transition. We use a linear switching term to interpolate the feature vectors of two atoms so make the feature vectors of the intermediate states by
\begin{align}
    I_{i-j,\epsilon-\epsilon'} = (1-\lambda)I_{i, \epsilon'} + \lambda I_{i, \epsilon},
\end{align}
where $\lambda$ is the switching parameter, $I_{i, \epsilon}$ and $I_{j, \epsilon'}$ are the feature vectors for atom $i$ which is present before the alchemical transition, and atom $j$ which is present after the alchemical transition. The elemental modes of atoms $i$ and $j$ are similarly interpolated before interacting with the feature vectors throught the $P$ and $Q$ matrices in equation 4. The interpolated feature vector is then fed into the network as normal to make predictions for the intermediate states. \\
\indent We used our network to run a molecular dynamics simulation with an alchemical transformation of an ethanol dimer into a water hexamer by slowly transitioning the carbon atoms of each ethanol into oxygen atoms. Figure \ref{fig:alchem} shows the atomization energies predicted by our network over this simulation. The total simulation time was 10 picoseconds (ps) with a 0.5 femtosecond (fs) time step. The first 2 ps occurred as purely the ethanol dimer allowing for some equilibration time. At 2 ps, the alchemical transition begun and was spread out over 3 ps. At 5 ps, the transition had completed leaving hydronium and hydroxyl ions along with water molecules which was followed by proton transfer at about 6 and 6.5 ps. We take particular note that during the transition time of 3 ps, no pathological behavior occurs, so we believe our network should be able to provide suitable alchemical free energies. We also note that because our implementation is in Google's TensorFlow package, then taking derivatives with respect to the switching parameter $\lambda$ is done by a single line of code and greatly simplifies a lot of the work needed for thermodynamic integration calculations.
\section{Discussion and Conclusions}
\indent Our work has presented a machine learned representation of atomic species, which is suitable to make improvements in several areas of computational and predictive chemistry. The atomic species representation is learned by compressing many physical properties into a smaller dimensional space using an auto-encoder. The compressed representation, which we have called the elemental modes, was shown to retain many of the periodic trends. \\ 
\indent We used the elemental modes to then show that they can perform well in tasks where we wish to rapidly screen materials such as elpasolites to help predict which structures may be stable for experimental researchers to pursue in further research. This same task can similarly be performed on other datasets of materials which all follow the same structural pattern and only differ in the atomic species. Prediction of other materials properties, such as band gaps should also be a trivial extension. Allowing more generalizations could be readily achievable as well. For example, allowing mixed species which occupy the D lattice site in elpasolites could be achieved by extending the feature vector of a material to allow for both species to exist. \\
\indent We also have shown the elemental modes to be useful in parameterizing neural network model chemistries. Previous works have been limited to predictions of molecules with only four different atomic species. Extending this was not straight forward since it would quickly cause issues with both computational efficiency and memory limitations of GPUs. The elemental modes allowed us to eliminate these problems and to improve the efficiency of training these neural networks. Further, we also showed the potential to simplify the process of making alchemical transformations. All of the code used in this work will be made available at www.github.com. Additionally the trained neural network for the elpasolite formation energy predictions will be available as well. \\
\indent Machine learning is becoming a well established method for making chemical predictions and reducing research costs. The vast size of chemical space is well beyond what can be explored by experiment and current computational methods alone. Machine learning algorithms show the promise to increase the rate at which we can find new candidate molecules and materials for many areas of important research by orders of magnitude. Representation of the candidate in the feature vector is certainly one of the most challenging and important parts of further research in this area. We believe that condensed representations such as the elemental modes will play an important role in improving much of the research to come in this blooming field.
\begin{acknowledgments}
The authors gratefully acknowledge Notre Dame's College of Science for startup funding, Oak Ridge national laboratory for a grant of supercomputer resources and Nvidia corporation.
\end{acknowledgments}

\bibliography{tensormol} 

\clearpage

\end{document}